\documentclass[aip,jcp,reprint,groupedaddress,showkeys,showpacs,longbibliography]{revtex4-1}

\usepackage{graphicx}
\usepackage{longtable}
\usepackage{url}
\usepackage{bm}
\usepackage{amsmath}
\usepackage{color}
\newcommand\kb{k_{\mathrm{B}}}

\begin{document}


\title{Mass-scaling replica-exchange molecular dynamics optimizes computational resources with simpler algorithm}


\author{Tetsuro NAGAI}
\email[]{tnagai@fc.ritsumei.ac.jp}
\author{Takuya TAKAHASHI}
\email[]{tkhs@sk.ritsumei.ac.jp}
\affiliation{Department of Bioinformatics, School of Life Sciences, Ritsumeikan University, Kusatsu, Siga 525-85077, Japan}


\date{\today}

\begin{abstract}
We develop a novel method of replica-exchange molecular dynamics (REMD) simulation, 
mass-scaling REMD (MSREMD) method, which improves trajectory accuracy at high temperatures, 
and thereby contributes to  numerical stability.  
In addition, the MSREMD method can also simplify a replica-exchange routine by eliminating velocity scaling. 
As a pilot system, a Lennard-Jones fluid is simulated with the new method. 
The results show that the MSREMD method improves the trajectory accuracy at high temperatures compared with the conventional REMD method.  
We analytically demonstrate that the MSREMD simulations can reproduce completely the same trajectories of the conventional REMD ones with shorter time steps at high temperatures in case of the Nos\'e-Hoover thermostats.  Accordingly, we can easily compare the computational costs of the REMD and MSREMD simulations.  
We conclude that the MSREMD method decreases the instability and optimizes the computational resources with simpler algorithm under the constant trajectory accuracy at all temperatures.
\end{abstract}


\maketitle

\section{Introduction}

Monte Carlo (MC) and molecular dynamics (MD)  simulations have been widely applied to many systems in the computational statistical physics field. 
However, the quasi-ergodicity problem, where simulations are prone to get trapped 
in states of energy local-minima, has been a great difficulty. 
In order to conquer this difficulty, generalized-ensemble algorithms 
have been developed and applied to many systems 
including spin systems and biomolecular systems (for reviews, see, e.g.,~Refs.~\onlinecite{Hansmann1999,Mitsutake2001,Sugita2002}).

Commonly practiced examples of the generalized-ensemble algorithms are  the multicanonical (MUCA) algorithm \cite{berg1991multicanonical,berg1992multicanonical}, 
the simulated tempering \cite{Lyubartsev1992,marinari1992simulated}, 
and the replica-exchange method (REM) \cite{hukushima1996exchange,Geyer1991} (it is also referred to as the parallel tempering). 
Closely related to MUCA are the Wang-Landau algorithm \cite{Wang2001a,Wang2001b} and metadynamics \cite{Laio2002}.
Also closely related to REM is the method in Ref.\ \onlinecite{Swendsen1986}, which is later detailed in Ref.~\onlinecite{wang2004replica}. 
The REM was first involved with MC simulations, and later the idea was also applied to MD simulations.  
The replica-exchange molecular dynamics (REMD)~\cite{sugita1999replica} method is the MD version of REM. 
Note that there are a number of attempts to generalize the REM and REMD, such as multi-dimensional extensions  
(see e.g.,~Refs.~\onlinecite{Sugita2000,Mitsutake2009multidimensional1,Mitsutake2009multidimensional2,Mitsutake2009MSTMREM,Nagai2012})
including the NPT ensemble \cite{Nishikawa2000,Okabe2001,Sugita2002,Paschek2004}
and the combination of the Tsallis statistics~\cite{Tsallis1988} with REM (see, e.g.,~Ref.~\onlinecite{kim2010}).

In this work, we particularly focus on a practical concern in application of the REMD method, in order to improve its efficiency. 
As the temperature increases, trajectory accuracy of simulations decreases and the simulations become numerically instable. 
Generally speaking, the shorter time step is necessary for the higher temperatures. 
However, this is not elegantly taken account of in previous applications.
The same time step is usually employed for all replicas. 
Some take risks of using a time step validated at low temperature for all replicas;  
others prudently employ a too short time step at low temperatures.   
For example, one of the authors has applied the REMD method to lipid bilayer systems with a coarse-grained model \cite{Marrink2004,Marrink2007}, and 
chose a shorter time step than that suggested in Ref.~\onlinecite{Marrink2004},   
to avoid the trajectory inaccuracy at high temperatures\cite{Nagai2012MolSim,Nagai2012JPSJ}. 
Such difficulty is also depicted in Ref.~\onlinecite{Han2010}, in which they employed a long time step for their force field, the Protein in Atomistic details coupled with 
Coarse-grained Environment (PACE), combing a united-atom and a coarse-grained force field. 
However, they tripled the mass of all proteins in the REMD simulations in order to avoid crashes at a high temperature.
We believe that there should be a number of preliminary simulations unpublished because such trajectory inaccuracy due to 
the usage of a long time step at high temperatures causes some numerical instability. 

One common practice for ensuring the accuracy and enhancing sampling efficiency is mass-scaling method which focused on increasing mass of hydrogen atom \cite{Pomes1990,Feenstra1999}. 
For details including the historical aspect, see Ref.~\onlinecite{Feenstra1999} and the references therein. 
The heavier mass enables one to use the larger time step with the same accuracy.  
This method is especially useful for quantum-mechanics/molecular-mechanics (QM/MM) simulations  
where covalent bond constraint algorithms such as SHAKE \cite{Ryckaert1977} RATTLE \cite{Andersen1983} are not suitable\cite{Zheng2009}, because chemical reactions involve bond forming and breaking. 
Note that there are other attempts of scaling masses to enhance the sampling efficiency \cite{Mao1991a,Mao1991b,Nguyen2010}.

In this article, using the idea mentioned above, we present a mass-scaling REMD (MSREMD) method, where masses of all particles are scaled
according to the reference temperature assigned to each replica. 
The heavier particles at the higher reference temperatures 
improve the trajectory accuracy of simulations. 
Furthermore, the MSREMD method does not require the velocity scaling necessary for the conventional REMD method and thereby the algorithm is simpler.

This article is organized as follows. In section II  we shall briefly review
the REMD method and introduce the MSREMD method.  
We prove that mass scaling in the equations of motion with the Nos\'e-Hoover thermostat \cite{Nose1984,Hoover1985}  is equivalent to the change of time step.   
Section III is devoted to Results and Discussion. 
We show that the MSREMD method restores the trajectory accuracy at the high temperatures. 
After we compare physical quantities obtained with the REMD and MSREMD methods, we contrast the estimated computational costs
of the REMD simulation with a long time step supposed to be verified at the lowest temperature, the REMD simulation with a short time step supposed to be validated at the highest temperature, and 
the MSREMD simulation.  
We conclude this paper in section IV with some outlook.

\section{Model and  Methods}
\subsection{Review of the REMD method\label{subsec:remd}}
 Before we present the MSREMD method,  we  shall simply review the REMD method.   
Readers who would like to know details about REM and REMD are referred to, e.g.,~Refs.~\onlinecite{sugita1999replica,hukushima1996exchange,frenkel2001understanding,berg2004book, Mori2010}. 
Because the MSREMD method follows the multi-dimensional REM formalization, we review the REMD method following Ref.~\onlinecite{Sugita2000}.  

We consider a system  consisting of $N$ particles, of which coordinate and momentum vectors are given by  
${q}=\{\bm{q}_1,\dots,\bm{q}_N\}$ and 
${p}=\{\bm{p}_1,\dots,\bm{p}_N\}$, respectively.
The velocity vector is denoted by $\dot{q}=\{\dot{\bm{q}}_1,\dots,\dot{\bm{q}}_N\}$, where the dot stands for the time derivative. 
The kinetic energy and the potential energy are denoted by $K$ and  $E$, respectively, with its total energy  $H=K+E$. 
In this work, we assume  $\bm{p}_k/m_k=\dot{\bm{q}}_k$ for $k=1,2,\dots, N$, where $m_k$ denotes the mass of the $k$th particle.  
The kinetic energy is therefore given by 
\begin{align}
K(\dot{q})&=\sum^N_k \frac{m_k\dot{\bm{q}}_k^2}{2}. 
\end{align} 
For the convenience in introducing the MSREMD method, we use $q$ and $\dot{q}$ instead of $q$ and $p$.
In the canonical ensemble at the reference temperature $T$, the state $\{{q},\dot{q}\}$ is weighted by the Boltzmann factor, 
\begin{align}
f_\text{NVT}(q,\dot{q}) \propto \exp [-H({q}, {\dot{q}})/\kb T] . 
\end{align} 

In the REMD method, $N_{\mathrm{rep}}$ copies of systems, namely replica 1, replica 2,$\dots$, and replica $N_{\mathrm{rep}}$ are simulated at the 
same condition except for the reference temperatures.  
Each replica is coupled to exclusively one of the $N_{\mathrm{rep}}$ different temperatures denoted by $T_1, T_2,\dots, T_{N_{\mathrm{rep}}}$.
For simplicity, we assume $T_1< T_2 < \dots< T_{N_\text{rep}}$.  
Every certain MD steps $N_{\mathrm{ex}}$, replicas attempt to exchange their reference temperatures. 
These exchanges of the temperatures cause replicas to perform a random walk in the temperature space, and this in turn induces a random walk in the energy space. 
The random walk of energy helps systems to overcome the energetic barriers.

In order to look into the REMD method further, we let replica $i$ be assigned to $T_l$. 
Due to its one-to-one correspondence, the replica index $i$ ($\in \{1,2,\dots,N_\text{rep}\}$) is given by
the permutation function of the temperature label $l$ ($\in \{1,2,\dots,N_\text{rep}\}$) and vice versa. 
We thus have
\begin{align} 
i=i(l)=l^{-1}(l),\\
l=l(i)=i^{-1}(i). 
\end{align} 

The replica-exchange attempts  are judged by Metropolis criterion \cite{metropolis1953equation}
, of which base is on the detailed balance condition;  the replica-exchange attempt between $T_l$ and $T_m$ is accepted at 
the probability of 
\begin{align}
\min[ 1, \exp (\Delta \beta \Delta E)],  \label{eq:Metropolis}
\end{align}
where $\Delta \beta =1/(\kb T_{l}) -1/(\kb T_{m})$ and $\Delta E = E_{i^\text{old}(l)}-E_{i^\text{old}(m)}$ 
with $\kb$ the Boltzmann constant and $E_{i^\text{old}(l)}$ the potential energy of the replica coupling to $T_l$ before the replica-exchange attempt. 
Although the Metropolis criterion for replica-exchange attempts is used,  
the Gibbs sampler \cite{chodera2011replica} and the Suwa-Todo method \cite{Suwa2010, Itoh2013PREMD} are also applicable. 

If the replica-exchange attempt is accepted, $i(l)$ will be reassigned:
\begin{align}
i^\mathrm{new} (l)=i^\mathrm{old}(m),\label{eq:label1}\\
i^\mathrm{new}(m) =i^\mathrm{old}(l),\label{eq:label2}  
\end{align} 
where the superscripts of old and new express before and after the accepted replica-exchange attempt, respectively.
 After the exchange, the velocities are requested to be uniformly scaled in the manner of
\begin{align}
\dot{q}'_{i^{\mathrm{new}(l  )}} &=\sqrt{\frac{T_l }{T_m}}\dot{q}_{i^{\mathrm{new}}(l  )}=\sqrt{\frac{T_l }{T_m}}\dot{q}_{i^{\mathrm{old}}(m )},\\
\dot{q}'_{i^{\mathrm{new}(m  )}} &=\sqrt{\frac{T_m }{T_l}}\dot{q}_{i^{\mathrm{new}}(m  )}=\sqrt{\frac{T_m }{T_l}}\dot{q}_{i^{\mathrm{old}}(l )},\\
\end{align} 
where $\dot{q}'_{i^\mathrm{new}(l)}$ stands for the velocity vector after the velocity scaling.  
The theoretical basis  for using such modified states is explicitly discussed in Ref.~\onlinecite{Nadler2007}.
The scaling of the velocities is required
because the kinetic energy distributions and reference temperatures are different from a replica to another.

When a constant-temperature MD simulation involve a deterministic thermostat with extra variables, a treatment particular to the thermostat is necessary to meet 
the detailed balance condition \cite{Mori2010}. 
The Nos\'e-Hoover thermostat \cite{Nose1984,Hoover1985} must be one of the most common thermostats for constant-temperature simulations, 
and we focus on the thermostat. 
Assuming ergodicity, the Nos\'e-Hoover thermostat with the reference temperature $T$ realizes the probability density function of $\{q, \dot{q}, \eta \}$:  
\begin{align}
&f_{\text{NH}} (q,\dot{q}, \eta)\\
&\propto \exp \left[ -\left( \sum_k^N \frac{m_k \dot{\bm{q}}_k^2}{2} +E(q)  +Q\eta^2/2 \right)/\kb T\right]\\
&=\exp \left[  -H(q,\dot{q})/\kb T  \right] \exp\left[ -Q\eta^2/2\kb T \right], 
\end{align}
where $\eta$ and $Q$  stand for the rate and mass of Nos\'e-Hoover thermostat, respectively. 
We again use the velocity instead of  the momentum for the later convenience. 
Integrating the probability density function with respect to $\eta$, one obtains the canonical distribution with regards to $\{q,\dot{q}\}$. 

Let us recall that the REMD method is based on the detailed balance condition. 
When the Nos\'e-Hoover thermostat is employed for REMD simulations, 
the detailed balance condition must be imposed considering  $\eta$ as well as $\{q,\dot{q}\}$.  
It is shown particularly  in Ref.~\onlinecite{Mori2010} that $\eta$ is also requested to be scaled similarly to the velocity.  
In practical, the detailed balance condition is 
fulfilled by setting $Q \propto T$ without scaling of $\eta$, 
which is implicitly practiced in, e.g.,~GROMACS software package\cite{Berendsen1995,Lindahl2001,VanDerSpoel2005,Pronk2013}.

We shall explore the probability density function of a system studied with the REMD method involving the Nos\'e-Hoover thermostats.  
We let the Greek letter $\xi_{i(l)}$  stand for  the state of the replica coupling to $T_l$: $\xi_{i(l)} =\{q_{i(l)},\dot{q}_{i(l)}, \eta_{i(l)} \}$.  
Because the replicas are noninteracting in the REMD method,  the state $X$ can be identified by specifying  all of the replicas: 
$X=\{{\xi_{i(1)}},\dots,\xi_{i(N_{\text{rep}})}\}$. 
Hence the REMD  simulation with the Nos\'e-Hoover thermostat has the probability density function of
\begin{align}
&\Pi_\text{REMD}(X)=\Pi_\text{REMD} (\xi_{i(1)},\dots,\xi_{i(N_\text{rep})}) = \prod_l f_{\text{NH}}(\xi_{i(l)})\notag 
\end{align}
\begin{align}
                  & \propto \prod_l\exp \left[ -\left( \sum_k^N\frac{m_{k} \dot{\bm{q}}_{k,i(l)}^2}{2} +E(q_{i(l)})  +Q\eta_{i(l)}^2/2 \right)/\kb T_l\right]. \label{eq:REMD}
\end{align}

\subsection{MSREMD method}
We now formalize the MSREMD method with the Nos\'e-Hoover thermostat. 
Note that the formalization is applicable to other thermostats such as Langevin\cite{Allen1989} and Andersen \cite{Andersen1980} ones
by eliminating the terms originating from the Nos\'e-Hoover thermostat.
Because the Hamiltonian is not necessarily identical 
among the replicas~\cite{Sugita2000},   
we choose the masses of all particles uniformly proportional to the reference temperature.  
Substituting $m_{k}=m_{0,k}T_l/T_1$ as well as $Q=Q_0\kb T_l$ for the Nos\'e-Hoover thermostat to Eq.~\ref{eq:REMD},   
we obtain
\begin{align}
&\Pi_\text{MSREMD} (X) \propto \notag\\
&\prod_l \exp \left[ -\left( \sum_k^N\frac{m_{0,k} \dot{\bm{q}}_{k,i(l)}^2}{2\kb T_1} +Q_0 \eta_{i(l)}^2/2\right) \right] \notag\\
&\times\exp\left[ -E({q}_{i(l)})/\kb T_l\right] , 
\end{align}
which signifies that the velocity vector $\dot{q}_{i(l)}$ shares the identical distribution among all of the reference temperatures and so does the Nos\'e-Hoover thermostat rate $\eta_{i(l)}$. 
Thus we can exchange the velocities and rates of the Nos\'e-Hoover thermostat without the care such as the Metropolis criterion or velocity scaling.
On the other hand, the coordinate vectors can be exchanged according to the probability given by Eq.~\ref{eq:Metropolis}, as in the REMD method. 
Moreover, the scaled masses enable one to restore the trajectory accuracy of simulation at high temperatures. 
Note that the exchange of the velocities instead of the momenta is crucial for this algorithm, because the momenta have different probability density functions among the reference temperatures. 

We explicitly show that the Metropolis criterion given by Eq.~\ref{eq:Metropolis} can be used for the replica-exchange attempts in the MSREMD simulation without the velocity scaling.  
Letting  $X^\text{old}$ and $X^\text{new}$ be 
\begin{align}
X^\text{old} = \left\{\xi_{i(1)},\dots,\xi_{i^\text{old}(l)},\dots,\xi_{i^\text{old}(m)},\dots,\xi_{i(N_\text{rep})}\right\}
\end{align}
 and
\begin{align}
X^\text{new}=\left\{\xi_{i(1)},\dots,\xi_{i^\text{new}(l)},\dots,\xi_{i^\text{new}(m)},\dots,\xi_{i(N_\text{rep})}\right\}, 
\end{align}
respectively, 
we look into the ratio of transition probabilities under the detailed balance condition between these two states, 
say $w(X^\text{old}\rightarrow X^\text{new})$ and $w(X^\text{new}\rightarrow X^\text{old})$.   
Using Eqs.~\ref{eq:label1} and \ref{eq:label2},  we obtain 
\begin{widetext}
\begin{align}
&\frac{w(X^\text{old}\rightarrow X^\text{new})}{w(X^\text{new}\rightarrow X^\text{old})}\notag\\
&=
\exp \left[ -\left( \sum_k^N\frac{m_{0,k} \dot{\bm{q}}_{k,i^\text{new}(l)}^2}{2\kb T_1} +Q_0 \eta_{i^\text{new}(l)}^2/2\right) \right] \exp\left[ -E({q}_{i^\text{new}(l)})/\kb T_l\right] \notag\\ 
&\times \exp \left[ -\left( \sum_k^N\frac{m_{0,k} \dot{\bm{q}}_{k,i^\text{new}(m)}^2}{2\kb T_1} +Q_0 \eta_{i^\text{new}(m)}^2/2\right) \right] \exp\left[ -E({q}_{i^\text{new}(m)})/\kb T_m\right] \notag\\
&\Biggl/\left(\exp \left[ -\left( \sum_k^N\frac{m_{0,k} \dot{\bm{q}}_{k,i^\text{old}(l)}^2}{2\kb T_1} +Q_0 \eta_{i^\text{old}(l)}^2/2\right) \right] \exp\left[ -E({q}_{i^\text{old}(l)})/\kb T_l\right] \right. \notag\\ 
&\left.\times \exp \left[ -\left( \sum_k^N\frac{m_{0,k} \dot{\bm{q}}_{k,i^\text{old}(m)}^2}{2\kb T_1} +Q_0 \eta_{i^\text{old}(m)}^2/2\right) \right] \exp\left[ -E({q}_{i^\text{old}(m)})/\kb T_m\right] \right)\notag\\
&=\exp\left[ ( E({q}_{i^\text{old}(l)})-E({q}_{i^\text{old}(m)}))(1/\kb T_l-1/\kb T_m)\right].  
\end{align}
\end{widetext}
In this way, we obtain the same term of $\Delta \beta\Delta T $ in Eq.~\ref{eq:Metropolis}.

To wrap up, MSREMD simulations with Nos\'e-Hoover thermostat can be performed as follows: 
(1) prepare $N_\text{rep}$ replicas with the masses of all of the particles and the Nos\'e-Hoover thermostats being in proportion to the reference temperature;  
(2) perform the $N_\text{rep}$ independent canonical MD simulations at each temperature; 
(3) exchange the replicas according to the probability given by Eq.~\ref{eq:Metropolis} without any velocity scaling; 
(4) go back to the step (2).

\subsection{Equations of motion of mass-scaled system\label{sec:scaling}}
We analytically demonstrate that mass scaling in the equation of motion of the  Nos\'e-Hoover thermostat  \cite{Nose1984,Hoover1985} is mathematically identical to changing the time step. 
The equations of motion are given by
\begin{align}
\dot{q}&=\frac{p}{m},\label{eq:NH1}\\
\dot{p}&=F(q)-\eta{p}, \label{eq:NH2}\\
\dot{\eta}&=\frac{p^2/m-3N\kb T}{Q},\label{eq:NH3}
\end{align}
where $F$ denotes force.  
For simplicity we set $m_1=m_2=\dots=m_N\equiv m$. 
We consider the transformations given by
\begin{align}
m&=\alpha m'\\
t&=\sqrt{\alpha} t' \label{eq:t}\\
Q&=\alpha Q'\\
q&=q'\label{eq:q}\\
\eta &=\frac{1}{\sqrt{\alpha}}\eta', 
\end{align}
where $\alpha$ denotes the scaling factor.
Substituting these transformations into Eqs.~\ref{eq:NH1}--\ref{eq:NH3},  we obtain  
\begin{align}
\frac{dq'}{dt'}&=\frac{p'}{m'},\\
\frac{dp'}{dt'}&=F(q')-\eta'{p'},\\
\frac{d\eta'}{dt'}&=\frac{p'^2/m'-3N\kb T}{Q'},
\end{align}
with $p'=p/\sqrt{\alpha}$. 
Therefore the time step $\Delta t$ of the system with the mass value $m$ correspond to 
the time step $\Delta t'=\Delta t/\sqrt{\alpha}$ of the system with the mass value $m'=m/\alpha$.  

We assume the primed quantities to be normal REMD ones at $T_l$, and the scaling factor $\alpha$ is $T_l/T_1$. 
The MSREMD simulation at $T_l$ with the time step $\Delta t$ generate 
the same time evolution of coordinate vector as the REMD simulation with the time step at $T_l$, 
\begin{align}  
\Delta t'(l) = \frac{\Delta t }{\sqrt{T_l/T_1}} .  
\end{align}  
On the other hand, the evolution of the momentum (or velocity) vector and rate of Nos\'e-Hoover thermostat is reproduced by using the scaling factor.

\subsection{Models} 
We employed a Lennard-Jones (LJ) fluid as a handful pilot system.  
The potential energy is given by 
\begin{align}
E&=\sum_{i<j} V_{ij}\\
V_{ij}&=4\epsilon \left[ \left( \frac{\sigma}{r_{ij}}\right)^{12} - \left(\frac{\sigma}{r_{ij}}\right)^{6} \right], 
\end{align}
where $\epsilon$ and $\sigma$ represent the value of the potential minimum and the diameter of particle, respectively, 
and $r_{ij}$ the distance between the $i$th and $j$th particles. 
Hereafter, we use reduced units; we set $\sigma=1$, $\epsilon=1$, and $\kb=1$.

\subsection{Numerical details}
The integrator was an in-house program. 
We used the time reversible integrator \cite{Martyna1996}. 
The choice of Suzuki-Trotter decomposition was made following Ref.~\onlinecite{Okumura2007}. 
This choice corresponds to Integrator 1 in Ref.~\onlinecite{Itoh2013DecOrder}. 
As a pseudo-random number generator for the replica-exchange routine, the Mersenne twister \cite{matsumoto1998mersenne} was employed. 

We performed both the REMD and MSREMD simulations. 
Three time steps $\Delta t$ were employed: $\Delta t=$0.002, 0.005, and 0.01 for each method. 
The total numbers of MD steps were $10^6$, $4\times 10^5$, and $2\times 10^5$ with the total time length fixed at 2000.  
The number of replica was eight ($N_\text{rep}=8$) and the reference temperatures were 1.000, 1.104, 1.219, 1.346, 1.486, 1.641, 1.812 and 2.000. 
The exchange acceptance rates ranged from 14\% to 19\% in all the simulations. 
Identical five-hundred LJ particles were placed ($N=500$) in a cube of which side was 8.55  in the reduced length unit,  
corresponding to the number density $\rho = 0.800$. 
In these thermal conditions, the LJ fluid is in the liquid phase~\cite{Okumura2000}. The periodic boundary condition was employed. 

The particle mass was set to unity for all of the replicas in the REMD simulations. 
On the other hand, the mass was given by $T_l/T_1$ for the replica coupling to $T_l$ in the MSREMD simulations.   
The mass of the Nos\'e-Hoover thermostat was set to $Q_0=10$ for the MSREMD and  REMD simulations.  
The replica-exchange attempts were made every $N_\text{ex}=10$ steps. 
At the MD step of $N_\text{ex}\times 2n$ for $n=1,2,3\dots$,  the replica-exchange attempts  were made 
between $T_1$ and $T_2$, between $T_3$ and $T_4$, between $T_5$ and $T_6$, and between $T_7$ and $T_8$.   
Correspondingly, at the MD step of $N_\text{ex}\times (2n-1)$,  the replica-exchange attempts were made 
between $T_2$ and $T_3$, between $T_4$ and $T_5$, and between $T_6$ and $T_7$.   
The LJ forces were simply truncated at 3 ($r_\text{c}=3$) in the reduced unit, and accordingly the LJ potential was shifted upward by $|V_{ij}(r_\text{c})|$ for $r_{ij}<r_\text{c}$.

\subsection{Evaluation of  Trajectory Accuracy}
We evaluated the simulation inaccuracy through the fluctuation of the conservation energy. 
The conservation energy of the Nos\'e-Hoover MD simulation with the reference temperature $T_\text{ref}$ is given by
\begin{align}
E_\text{cons}(t)&=\notag\\
\sum_k^N & \frac{\bm{p}_k^2}{2m_k} +E(q) +\frac{1}{2}Q\eta^2 +3N\kb T_\text{ref} \int ^t_0 \eta(t') dt'.
\end{align}
Practically, this quantity fluctuates reflecting numerical errors. 
Thus the sum of absolute fluctuation per unit time 
is used for the evaluation of the trajectory accuracy:  
\begin{align}
\Delta E_\text{cons} \equiv %
\frac{1}{\Delta t N_\text{step} }\sum_{i=1}^{N_\text{step}} |E_\text{cons}(i\Delta t)-E_\text{cons}((i-1)\Delta t)|, 
\end{align}
where $N_\text{step}$ is the number of MD steps. 
Note that the way of evaluation is similar to Eq.~74 in Ref.~\onlinecite{Martyna1996}, which is given by
\begin{align}
\frac{1}{N_\text{rep}}\sum_{i=1}^{N_\text{step}} \left|\frac{E_\text{cons}(i\Delta t)-E_\text{cons}(0)}{E_\text{cons}(0)}\right|. 
\end{align}
We used the deviations in the conservation energy between the successive time steps because $E_\text{cons}(0)$ changes due to the accepted replica-exchange attempts. 
We checked the relationship between $\log \Delta t$ and $\log \Delta E_\text{cons}$ and confirmed that the gradient values were 2.0 and almost agree with those in Ref.~\onlinecite{Martyna1996}. 
 
\section{Results and Discussion}
\begin{figure}[bht]
        \begin{center}
            \includegraphics[width=5.5cm, clip , angle = 270]{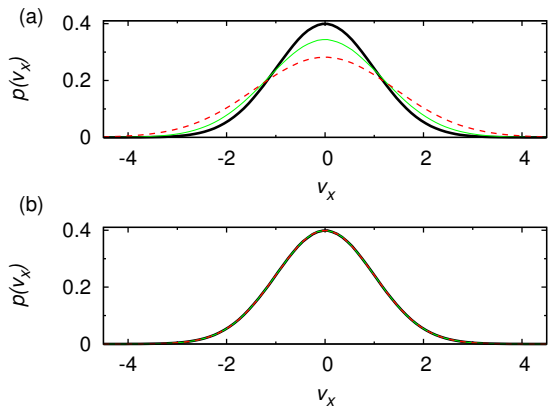}
            \caption{Probability density functions of $v_x$, $p(v_x)$, for (a) REMD and (b) MSREMD simulations. 
             Black thick, green thin, and red dashed lines represent the functions for $T_1$, $T_4$ and $T_8$, respectively. 
			Those at other temperatures are suppressed to aid the eye.  Lines completely overlap in MSREMD simulation [see (b)]. }
            \label{fig:v-dist}
        \end{center}
\end{figure}

Figure \ref{fig:v-dist} shows the probability density functions of the velocity $x$-component, $v_x$, obtained with the REMD and MSREMD simulations for $\Delta t = 0.01$.   
We took account of all of the particles. 
The velocities at the  half time steps were used in the calculation of the kinetic energy  according to the recommendation in Ref.~\onlinecite{Itoh2013DecOrder}. 
The probability density functions were different among the temperatures in the  REMD simulation, whereas 
those obtained in the MSREMD simulations were the same among the temperatures. These results show that  the proper scaling of mass enables one to produce 
the same probability density function of velocity.  

\begin{figure}[bht]
        \begin{center}
            \includegraphics[width=5.5cm, clip , angle = 270]{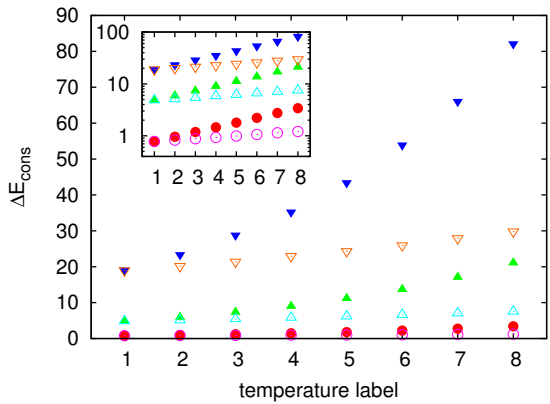}
            \caption{Trajectory accuracy $\Delta E_\text{cons}$ plotted against temperature label.  
                        Red closed circles, green closed triangles and blue closed inversed triangles represent the results of normal REMD simulations for $\Delta t=$0.002, 0.005 and 0.01, respectively. 
                        Magenta open circles, cyan open triangles and orange open inversed triangles show the results of MSREMD simulations for $\Delta t=$0.002, 0.005 and 0.01, respectively.  
                        Error bars are so small that they are suppressed to aid the eye. The inset is the same plot with the logarithm ordinate.              }
            \label{fig:dtEconv}
        \end{center}
\end{figure}

Figure \ref{fig:dtEconv} shows the trajectory inaccuracy of the simulations, as measured by 
$\Delta E_\text{cons}$, with an inset of the logarithmic ordinate. 
The REMD simulations become more inaccurate as the temperature increases. 
In contrast, the trajectory accuracy at high temperatures obtained with the MSREMD simulations is of the same level as the low temperatures for each time step.  
The slight increases in the trajectory inaccuracy with regards to the temperature  in the MSREMD simulation could be attributed to the steeper potential surfaces faced at higher potential energy values.   
Therefore, 
the trajectory inaccuracy at high temperature can be substantially reduced by the MSREMD method, 
which infers that one can perform more numerically stable simulations with the MSREMD method.  

\begin{figure}[bt]
        \begin{center}
              \includegraphics[width=5.5cm, clip , angle = 270]{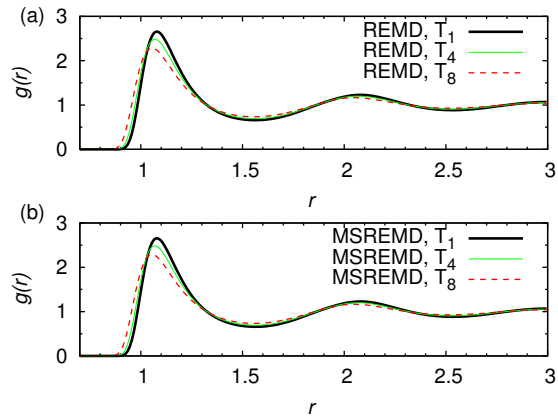}
            \caption{Radial distribution functions $g$ for (a) the normal REMD and (b) MSREMD simulations, as a function of distance $r$ between particles. 
                         Black thick, green thin, and red dashed lines represent replica 1, 4 and 8, respectively. 
			The other replicas  are suppressed to aid the eye.  }
            \label{fig:rdf}
        \end{center}
\end{figure}

Table \ref{table} shows the average potential and kinetic energies obtained with the REMD and MSREMD simulations for all of the reference temperatures. 
The errors in Table \ref{table} were obtained by using the jackknife method \cite{efron1982jackknife,Flyvbjerg1989,berg2004book} with twenty bins.  
Figure \ref{fig:rdf} shows the radial distribution function for $T_1$, $T_4$ and $T_8$ obtained with the two methods. 
These results show the very good agreement between the two methods, which is very natural consequences. 
All of the coordinate-related quantities must be the same between the two methods because the changes of masses do not affect the configurational  partition function. 
In addition, the same amount of the kinetic energy ought to be distributed to each degree of freedom at the same temperature regardless of the weight of particles,  
due to the equipartition theorem in the classical statistical physics. 
Note that we did not find appreciable difference in the sizes of errors between the two methods (see Table \ref{table}). 
\begin{table*}[th]
\caption{Average kinetic energy $\left<K\right>$ and average potential energy $\left<E\right>$ obtained with the two methods at  all the reference temperatures. 
 Kinetic energy was calculated with the velocities at half time steps \cite{Itoh2013DecOrder}.
 Errors were evaluated with the jackknife method. 
 Exact average kinetic energy obtained by $\frac{3}{2}NT$ is also given. }
\label{table}
\begin{tabular}{lcccccc}
\hline\hline
  & &\multicolumn{2}{c}{REMD} && \multicolumn{2}{c}{MSREMD} \\\cline{3-4}\cline{6-7}  
$T_l$ &$\frac{3}{2}NT_l$ &$\left<K \right>$&$\left<E \right>$  & & $\left<K \right>$&$\left<E \right>$\\\hline
1.000  &750.00   &750.02  $\pm$  0.05  & -2519.3  $\pm$ 0.1   && 750.11 $\pm$ 0.05 & -2519.7 $\pm$  0.1\\ 
1.104  &828.00   &827.96  $\pm$  0.08  & -2474.1  $\pm$ 0.2 &&  828.0    $\pm$ 0.1  &  -2474.5  $\pm$ 0.2\\
1.219  &914.25   &914.34   $\pm$ 0.09  & -2425.8  $\pm$ 0.2 &&  914.1    $\pm$ 0.1  &  -2426.1  $\pm$ 0.2\\ 
1.346  &1009.5   &1009.45  $\pm$ 0.09 & -2374.2 $\pm$ 0.2   &&  1009.5  $\pm$ 0.1  & -2373.9 $\pm$ 0.2\\ 
1.486  &1114.50 & 1114.5  $\pm$ 0.1    & -2319.2  $\pm$ 0.2 &&  1114.7  $\pm$ 0.1 &   -2318.5 $\pm$   0.2\\ 
1.641  &1230.75 & 1230.6  $\pm$ 0.1    & -2259.3  $\pm$0.2  &&  1230.8  $\pm$ 0.1 &   -2258.9 $\pm$   0.2\\ 
1.812  &1359.0   &1359.1   $\pm$ 0.1   & -2195.8 $\pm$ 0.2   &&  1358.9 $\pm$  0.1  &  -2195.3  $\pm$  0.3 \\ 
2.000  &1500.0   &1500.0   $\pm$ 0.1   & -2128.1 $\pm$  0.2  &&  1500.0 $\pm$  0.1 &   -2128.3 $\pm$ 0.3\\\hline\hline
\end{tabular}
\end{table*}

We compare the computational costs of the three simulations:
the long-time-step REMD (LTS-REMD) simulation of which time step $\Delta t_\text{L}$ is validated at the lowest temperature; 
the short-time-step REMD (STS-REMD) simulation with the time step  $\Delta t_\text{S} = \sqrt{T_1/T_{N_\text{rep}}}\Delta t_\text{L}$ prudently validated at the highest temperature; 
and the MSREMD simulation with the time step $\Delta t_\text{L}$. 

As a measure of the computational cost, we calculate the efficiency ratio of the STS-REMD simulation to the LTS-REMD simulation, $f_\text{S}$, as follows.  
Because the time step is given by $\Delta t_\text{S} = \sqrt{T_1/T_{N_\text{rep}}}\Delta t_\text{L}$, 
the trajectory obtained with the STS-REMD simulation is $\sqrt{{T_1}/{T_{N_\text{rep}}}}$ times as long as that obtained with the LTS-REMD simulation, which yields the efficiency ratio, 
\begin{align}
f_\text{S}(Z)=\sqrt{\frac{T_1}{T_{N\text{rep}}}}=\frac{1}{\sqrt{Z}}  <1 ,  
\end{align} 
where a temperature ratio $Z$ is ${T_{N_\text{rep}}}/{T_{1}}$. 
The efficiency ratio of the LTS-REMD simulation to the LTS-REMD simulation, $f_\text{L}$ is obviously unity.

\begin{figure}[!bht]
        \begin{center}
            \includegraphics[height=5.5cm, clip ]{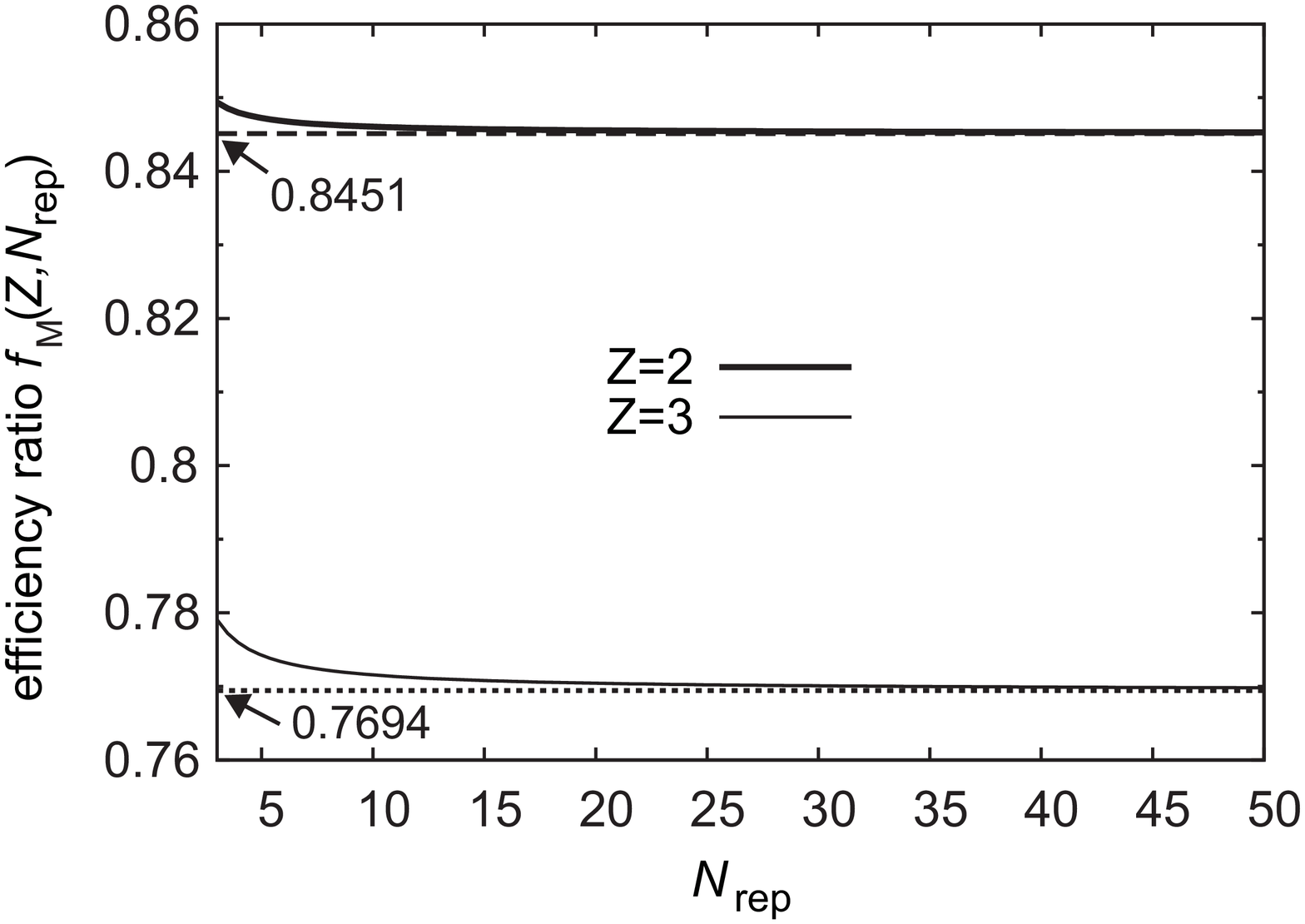}
            \caption{ Efficiency  ratio $f_\text{M}$  plotted  against the number of replicas, $N_\text{rep}$.
                         Thick solid, and thin solid lines represent the efficiency ratios $f_\text{M}(2,N_\text{rep})$ and $f_\text{M}(3,N_\text{rep})$, respectively. 
                         Dashed and dotted lines approximately
                          correspond to the values of $f_\text{M}(2,\infty)$ and $f_\text{M}(3,\infty)$, respectively.  } 
            \label{fig:slowdown}
            \includegraphics[height=5.5cm, clip]{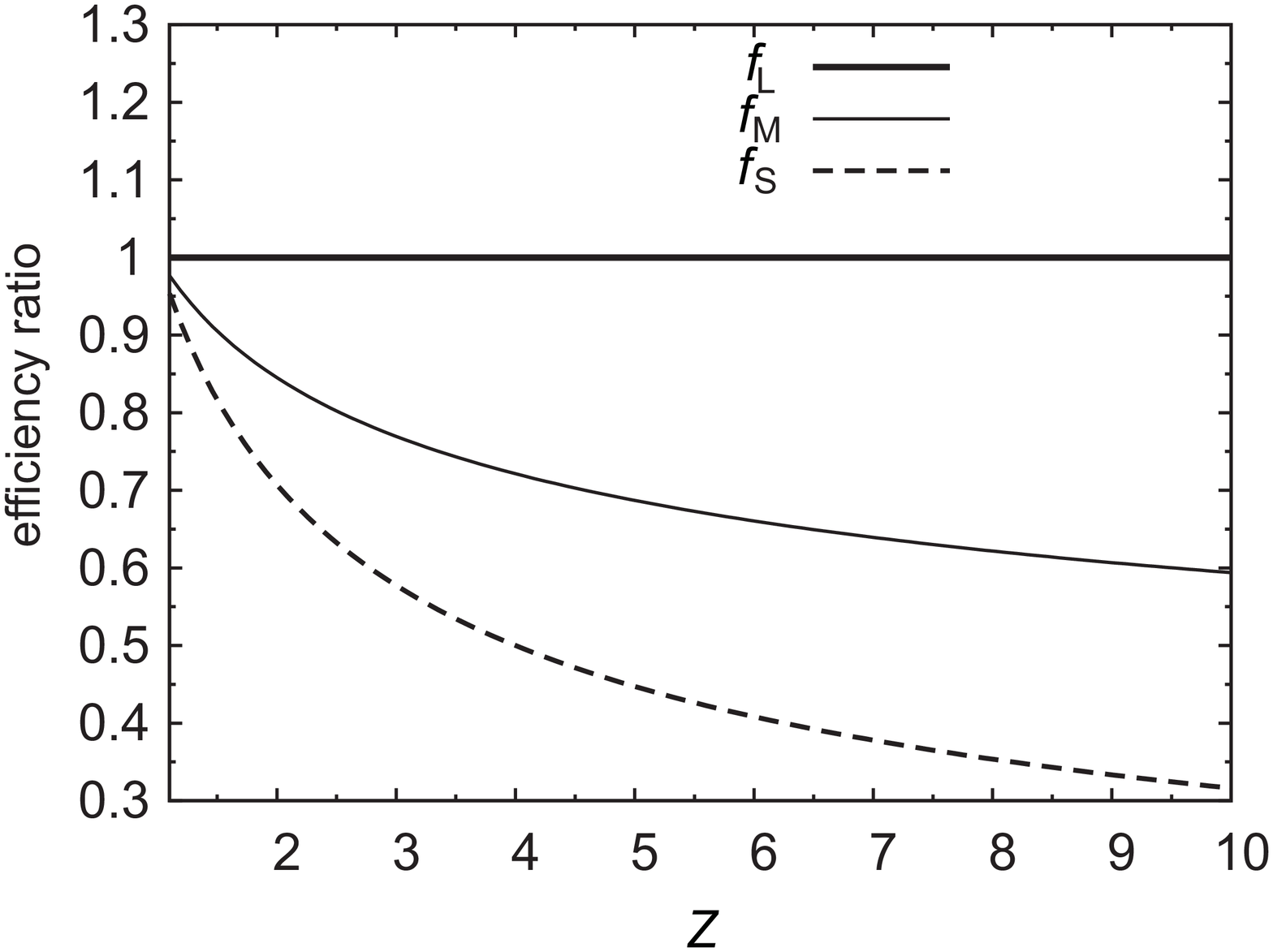}
            \caption{Efficiency ratios $f_\text{L}$, $f_\text{M}$, and $f_\text{S}$ as functions of the temperature ratio $Z$.  
                         Thick solid, thin solid, and dashed lines represent $f_\text{L}(Z)$, $f_\text{M}(Z,\infty)$ and 
                         $f_\text{S}(Z)$, respectively.
                          } 
            \label{fig:slowdown2}
        \end{center}
\end{figure}

Owing to the correspondence shown in section \ref{sec:scaling}, the trajectory belonging to $T_l$ in the MSREMD simulation is
$\sqrt{T_1/T_l}$ ($\leq 1$)  times as long  as that in the LTS-REMD simulation.
Consequently, the efficiency ratio of the MSREMD simulation to the LTS-REMD simulation is given by  
\begin{align}
\frac{1}{N_\text{rep}}\sum_{l=1}^{N_\text{rep}} \sqrt{\frac{T_1}{T_l}}. 
\end{align}
Because the temperatures are usually given according to a geometric series,  
the efficiency ratio $f_\text{M}(Z,N_{\text{rep}})$ turns out to be
\begin{align}
f_\text{M} (Z,N_{\text{rep}})=\ \frac{1}{N_\text{rep}}
\frac{
1-\left(\frac{1}{\sqrt{Z}}\right)^{\frac{N_\text{rep}}{N_\text{rep}-1}}
}{
1-\left(\frac{1}{\sqrt{Z}}\right)^{\frac{1}{N_\text{rep}-1}}
} . 
\end{align}
For $N_\text{rep}\rightarrow\infty$,  $f_\text{M}$ is given by
\begin{align}
f _\text{M}(Z,\infty)\equiv \lim_{N_\text{rep}\rightarrow \infty} f_\text{M} (Z,N_\text{rep})= \frac{1-\left(\frac{1}{\sqrt{Z}}\right)}{\ln\sqrt{Z}}. 
\end{align}
The values of $f_\text{M}(Z,\infty)$ are $0.8451\dots$ and $0.7694\dots$ for $Z=2$ and $3$, respectively.   
Note that $Z=3$ corresponds to $T_1=300$~K and $T_{N_\text{rep}}=900$~K, which should represent the popular 
application of the REMD method to all-atom simulations.  
The limit value of $f_\text{M}(Z,\infty)$  is the lower bound as is illustrated in Figure \ref{fig:slowdown}, which shows how  $f_\text{M}$  converges for $Z=2$ and $Z=3$ as $N_\text{rep}$ tends to infinity. 
The convergence is fast and we use the limit values as the value of efficiency ratio of the MSREMD simulation to the LTS-REMD simulation. 

Figure \ref{fig:slowdown2} illustrates $f_\text{L}$, $f_\text{S}$, and $f_\text{M}$ for $N_\text{rep}=\infty$ as functions of $Z$. 
The efficiency of the STS-REMD and MSREMD simulations decrease as $Z$ increases. 
The MSREMD simulation is more efficient than the STS-REMD simulation ($f_\text{S}<f_\text{M}$). 
The efficiency ratio of the MSREMD simulation to the STS-REMD simulation is given by  
\begin{align}
\frac{f_\text{M}}{f_\text{S}}=\frac{\sqrt{Z}-1}{\ln\sqrt{Z}}.
\end{align}
The values of ${f_\text{M}}/{f_\text{S}}$ for $Z=2$ and $Z=3$ are $1.195\dots$ and $1.333\dots$, respectively. 
Therefore the MSREMD simulation is 20\% to 30\% more efficient than the STS-REMD simulation.

\section{Conclusions}
We introduced the MSREMD method, where we scale the masses of all the particles uniformly proportional to the reference temperatures. 
We analytically showed that the scaling of mass in the equations of motion with the Nos\'e-Hoover thermostat corresponds to the scaling of time step.  
The larger masses at the higher reference temperatures help one restore the trajectory accuracy at the high temperatures, 
which infers that  more stable simulations are feasible. 
Moreover, the identicalness of  the velocity distributions realized by the MSREMD method enables one to exchange the replicas without velocity scaling, 
and thereby the replica-exchange routine is simpler. 
Because we only manipulate mass values in the MSREMD method, 
the  coordinate-related quantities such as the radial distribution function and the average potential energy are identical to those obtained with the REMD method. 
The kinetic energy distributions and the heat capacities are also identical between the two methods.  

We evaluated the efficiency ratios of the STS-REMD and MSREMD simulations to the LTS-REMD simulation. 
The MSREMD simulation should typically use 20\% to 30\% more resources than the LTS-REMD simulation with the potentially risky time step validated at the  lowest temperature. 
On the other hand, the MSREMD  simulation typically uses 20\% to 30\% less computational resources than the STS-REMD simulation with the short time step validated at high temperatures.
The MSREMD method therefore balances the trajectory accuracy and the computational cost by effectively adjusting the time steps  according to the reference temperatures. 

One interesting extension of the MSREMD method for biomolecules would be to change the way of scaling according to the atom species as well. 
Such an extension would be useful for e.g.~the QM/MM simulations, where the covalent bond constraint algorithms are not suitable.  
We also expect that the new method works well with coarse-grained models.   
Whereas we particularly focused on the NVT ensemble, 
the rigorous formalization and evaluation of the MSREMD method with other thermostats or other ensembles are our interesting future task.

\begin{acknowledgments}
Some of  computations  were performed at the Research Center for Computational Science, Okazaki, Japan. 
This work was, in part, supported by  Grant-in-Aid for Young Scientists (B) under Grant No.~26790083. 
TT gratefully acknowledges support by Grant-in-Aid for Scientific Research (C) under Grant No.~25440065 and 
Grant-in-Aid for Scientific Research on Innovative Areas under Grant No.~20118003.

\end{acknowledgments}

%

\end{document}